\documentclass[showpacs,aps,pra,floatfix,reprint,superscriptaddress,footinbib,citeautoscript,nofootinbib]{revtex4-2}
\usepackage{amssymb,amsfonts,amsmath}
\usepackage{graphicx}
\usepackage{verbatim}
\usepackage[utf8]{inputenc}
\usepackage[T1]{fontenc}
\usepackage{soul,xcolor,colortbl}
\usepackage{multirow}
\usepackage{bm}
\usepackage{array}
\usepackage{chemmacros}
\usepackage{color}
\usepackage{hyperref} \hypersetup{colorlinks=true,citecolor=blue,linkcolor=blue,urlcolor=blue}
\usepackage{mdframed}
\usepackage{longtable}
\usepackage{enumitem}
\usepackage{booktabs}
\usepackage{lipsum}
\usepackage{orcidlink}
\usepackage{paralist}
\usepackage{listings}

\newcolumntype{L}[1]{>{\raggedright\arraybackslash}p{#1} }
\newcolumntype{C}[1]{>{\centering \arraybackslash}p{#1} }
\newcolumntype{R}[1]{>{\raggedleft \arraybackslash}p{#1} }

\DeclareSIUnit\inch{inches}

\newcommand{\asubsection}[1]{\vspace{5mm}\noindent \textbf{#1}\\}
\newcommand{\aasubsection}[1]{\vspace{5mm}\noindent \textbf{#1}\ \ }

\newlist{noteize}{itemize}{4}
\setlist[noteize]{label=\textcolor{blue}{\textbullet}, font=\footnotesize, noitemsep, align=parleft, labelwidth=0.5em, leftmargin=1em}

\def\AFLOW{{\small AFLOW}}
\def\AFLOWF{{\small AFLOW4}}
\def\AFLOWorg{{\tt\small aflow.org}}
\def\DFT{{\small DFT}}
\def\HPC{{\small HPC}}

\def\JSON{{\small JSON}}
\def\VASP{{\small VASP}}

\def\AUID{{\small AUID}}
\def\DEED{{\small DEED}}
\def\CCE{{\small CCE}}
\def\LDA{{\small LDA}}
\def\SCAN{{\small SCAN}}
\def\PBE{{\small PBE}}
\def\CHULL{{\small CHULL}}
\def\POCC{{\small POCC}}
\def\CPOCC{{\small cPOCC}}
\def\years{25}

\definecolor{codegreen}{rgb}{0,0.6,0}
\definecolor{codegray}{rgb}{0.5,0.5,0.5}
\definecolor{codepurple}{rgb}{0.58,0,0.82}

\lstdefinestyle{aflow}
{
    frame=leftline,
    breaklines=true,
    backgroundcolor=\color{gray!10!white},
    basicstyle=\scriptsize\color{black}\ttfamily
}

\lstdefinelanguage{aflowBash}
{
    keywords=[1]{
        aflow,tachyon,cd,bash
    },
    keywordstyle=[1]\color{codepurple},
    keywords=[2]{
        render,prototype,proto,params,
        alloy,chull,aflowlibpath,export,
        soliquidy,auid,o,format,auto_skylight,
        fullshade,numthreads,pocc_params,
        pocc_count_unique_fast,cce
    },
    keywordstyle=[2]\color{codegreen},
    keywords=[3]{
        ALLOY, STRUCT_FILE,PATH
    },
    keywordstyle=[3]\color{codegray},,
    sensitive=true
}

\setcitestyle{square}

\makeatletter \renewcommand\frontmatter@abstractwidth{\dimexpr\textwidth\relax} \makeatother

\def\MEMS{Department of Mechanical Engineering and Materials Science, Duke University, Durham, NC 27708, USA}
\def\CEM{Center for Extreme Materials, Duke University, Durham, NC 27708, USA}
\def\CNRMODENA{\footnotesize CNR-NANO Research Center S3, 41125  Modena, Italy}
\def\ZENTRUM{\footnotesize Institute of Ion Beam Physics and Materials Research, Helmholtz-Zentrum Dresden-Rossendorf, 01328 Dresden, Germany}
\def\TCTUD{\footnotesize Theoretische Chemie, Technische Universit\"at Dresden, 01062 Dresden, Germany}

\begin{document}
\title{AFLOW4: heading toward disorder}
\author{Simon~Divilov\,\orcidlink{0000-0002-4185-6150}$^\dagger$}\affiliation{\MEMS}\affiliation{\CEM}
\author{Hagen~Eckert\,\orcidlink{0000-0003-4771-1435}$^\dagger$}\affiliation{\MEMS}\affiliation{\CEM}
\author{Scott~D.~Thiel\,\orcidlink{0000-0002-9947-0277}}\affiliation{\MEMS}\affiliation{\CEM}
\author{Sean~D.~Griesemer\,\orcidlink{0000-0001-5531-0725}}\affiliation{\MEMS}\affiliation{\CEM}
\author{Rico~Friedrich\,\orcidlink{0000-0002-4066-3840}}\affiliation{\CEM}\affiliation{\ZENTRUM}\affiliation{\TCTUD}
\author{Nicholas~H.~Anderson\,\orcidlink{0009-0008-4851-6731}}\affiliation{\MEMS}\affiliation{\CEM}
\author{Michael~J.~Mehl\,\orcidlink{0000-0001-9402-6591}}\affiliation{\MEMS}\affiliation{\CEM}
\author{David~Hicks\,\orcidlink{0000-0001-5813-6785}}\affiliation{\CEM}
\author{Marco~Esters\,\orcidlink{0000-0002-8793-2200}}\affiliation{\CEM}
\author{Nico~Hotz\,\orcidlink{0009-0008-2469-2693}}\affiliation{\MEMS}\affiliation{\CEM}
\author{Xiomara~Campilongo\,\orcidlink{0000-0001-6123-8117}}\affiliation{\CEM}
\author{Arrigo~Calzolari\,\orcidlink{0000-0002-0244-7717}}\affiliation{\CNRMODENA}\affiliation{\MEMS}\affiliation{\CEM}
\author{Stefano~Curtarolo\,\orcidlink{0000-0003-0570-8238}}\email[]{stefano@duke.edu}\affiliation{\MEMS}\affiliation{\CEM}

\date{\today}

\begin{abstract}
 \noindent
\AFLOWF\ is the latest iteration of the \AFLOW\ toolkit, specifically tailored to study high-entropy disordered materials.
This upgrade includes innovative features like the Soliquidy module, based on the Euclidean transport cost between disordered and ordered material states.
\AFLOWF\ can calculate dielectric functions to understand optical and electronic properties of disordered ceramics.
The newly introduced human-readable data export feature ensures the uncomplicated incorporation of \AFLOWF\ in diverse automated workflows.
Features relevant to high-entropy research, like prototype identification, partial occupation method, convex hull calculation, and enthalpy corrections based on local atomic environments, have been improved and exhibit substantial speed-up.
Together, these enhancements represent a step forward for \AFLOW\ as a valuable tool for research of high-entropy materials.
\end{abstract}

\maketitle

\section*{Introduction}
The \underline{A}utomatic-\underline{Flow} (\AFLOW) framework is a class of interconnected software tools designed for high-throughput materials discovery, leveraging both first-principles quantum mechanical calculations using density functional theory (\DFT), and data informatics as a screening strategy~\cite{aflowPAPER}.
\AFLOW\ has been in development for over {\years} years and comprises over \num{500000} lines of \texttt{C++} code.
Altogether, it now contains 15 modules or tools valuable for materials characterization and exploration.

In tandem, \AFLOWorg, a web ecosystem of applications mirroring the functionality of AFLOW, and databases following the FAIR (\underline{F}indable, \underline{A}ccessible, \underline{I}nteroperable, and \underline{R}eusable) principles~\cite{Wilkinson_FAIR_SciData_2016} provide both graphical and programmatic methods to access over 3.5 million material entries~\cite{aflowlib,curtarolo:art190}.
The databases also contain over \num{2000} crystallographic prototypes~\cite{curtarolo:art210}, used to uniquely classify crystal structures.
For users who require only a subset of \AFLOW\ features, such as to create convex hulls for thermodynamic analysis, or who are unfamiliar with compilation and command line interfaces, \AFLOWorg\ is an ideal choice to find the necessary information for their research.

More recently, we have been actively optimizing and developing original modules in \AFLOW\ to facilitate the discovery of novel chemically and structurally disordered high-entropy materials which often exhibit attractive and exotic properties~\cite{Cantor_ENTROPY_2014,HEA_review,curtarolo:art163,curtarolo:art187,Hsu_NRC_2024}.
Therefore, to handle the huge number of possible crystal structures due to the vast compositional space has required upgrading our high-throughput techniques.
This means that existing code needed refactoring to take advantage of modern \texttt{C++17} features that did not exist when \AFLOW\ was first designed at MIT during 1999-2003~\cite{curtarolo:mit_thesis}.

As we mark its {\years}th anniversary, we release \AFLOW's fourth iteration, \AFLOWF, specifically designed with high-entropy systems in mind.
We have improved existing routines which are used to generate an extensive amount of data, ranging from electronic, elastic, optical, vibrational, and thermodynamic.
In addition, we have added new workflows to extend the functionality of \AFLOW\ in calculating new kinds of systems and analyzing the produced raw output.
Overall, we have updated almost all parts of \AFLOW, including how users can acquire and install the code and how it is documented.
The upgrades and optimizations were done with portability in mind, allowing \AFLOW\ to run efficiently on a wide variety of computer systems, ranging from personal laptops to high-performance computing (\HPC) systems.

The article is structured into two sections: ``\textit{\AFLOWF\ for high-entropy Materials}'' and ``\textit{\AFLOWF\ Development Update}''. In the first section, we provide an overview of significant techniques that have recently been implemented in \AFLOW. The second one introduces the fundamental concepts for downloading, building, and documenting the \AFLOW\ code.
The new or updated features in \AFLOW\ include calculation of Soliquidy, calculation of the dielectric function, human-readable data export, prototype identification, the partial occupation method, calculation of the convex hull and, enthalpy corrections based on the local atomic environment.

\section*{AFLOW4 for High-Entropy Materials}
\label{sec:usage}
High-entropy materials overcome several limitations of conventional materials using chemical and structural disorder.
They find applications in thermal barrier protection, wear- and corrosion-resistant coatings, thermoelectrics, batteries, and catalysts~\cite{Yeh_JMR_2018,Gao_JMR_2018}.
Consequently, there is a pressing need for computational tools capable of handling the enormous composition space these systems present.

In this release, we also introduce two new modules:
(i) the Soliquidy module that calculates a geometric descriptor that numerically captures the Euclidean transport cost between the translationally disordered versus ordered states of a material~\cite{curtarolo:art211};
(ii) the Export module that generates machine-readable data files of database entries to be used, for example, as reference data for parameterizing machine-learning potentials.

\asubsection{Soliquidy}
In this update to the \AFLOW\ code base, we introduce the Soliquidy module~\cite{curtarolo:art211}, a novel functionality designed to quantify the ease with which materials transition from disordered states, such as liquids or vapors, to ordered crystalline structures.
Soliquidy frames this transition as an optimal transport problem.
This mathematical approach seeks the most efficient way to rearrange atoms from their initial disordered positions to their final ordered ones, minimizing the \textit{cost} or effort required for this movement.
Our implementation uses weight-optimized Voronoi cells to solve the optimal transport problem.
Particularly, we employ the Kitagawa, M\`erigot, and Thibert ({\small KMT}) algorithm~\cite{kitagawaConvergenceNewtonAlgorithm2019}.

\begin{figure}[ht!]
  \includegraphics[width=0.5\textwidth]{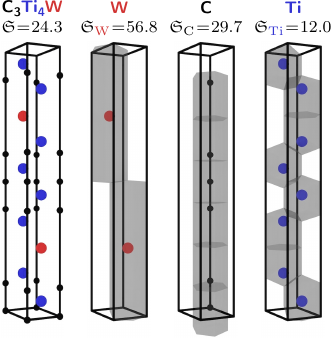}
  \caption{\small \textbf{Application of Soliquidy ($\mathfrak{S}$)} to \ch{C3Ti4W}
    (\href{https://aflow.org/material/?id=03cb24fa8c754e9d}{aflow:03cb24fa8c754e9d})
    in the spacegroup $P6_3/mmc$ (\#194). The volumes highlighted in
    gray for each element are the equal sized Voronoi cells, that form
    the basis of the Soliquidy descriptor. Reproduced from Ref.~\onlinecite{curtarolo:art211} (CC BY-NC-ND 4.0).}
  \label{fig:soliquidy}
\end{figure}

Lower values of Soliquidy indicate easier transitions, favoring crystallization, while higher values suggest greater movement is needed, predisposing materials towards glass formation.
This module gives users access to this predictive measure that aids in material design and understanding phase transitions.
It can either be invoked for a specific entry in the \texttt{aflow.org} database using its unique identifier (\AUID).
For example, \texttt{03cb24fa8c754e9d} will give you the information for a \ch{C3Ti4W} structure, shown in Fig.~\ref{fig:soliquidy}.
\begin{lstlisting}[style=aflow,language=aflowBash]
aflow --soliquidy --auid=03cb24fa8c754e9d
\end{lstlisting}
Alternatively, it is also possible to provide a structure file directly on the standard input.
\begin{lstlisting}[style=aflow,language=aflowBash]
aflow --soliquidy < STRUCT_FILE
\end{lstlisting}
The output generated on the command line for both examples is a \JSON\ representation that includes the Soliquidy result of the structure under the key \texttt{value} and broken down for each element under the key \texttt{species}.

Additionally, \AFLOW\ assists in generating graphical representations of the Voronoi cells that form the foundation of the optimal transport problem.
Using the \texttt{render} keyword, three distinct variations are generated for each element within the structure in the output folder:
(i)~an \texttt{html} file, incorporating a 3D scene in the \texttt{x3D} format, enabling the users to swiftly inspect the results in any contemporary web browser;
(ii) a \texttt{ty} file, which can be used with the open-source raytracer \texttt{tachyon}~\cite{stoneEfficientLibraryParallel1998} to produce high-quality images; and (iii) folders containing the necessary data to generate animations using \texttt{tachyon} and \texttt{ffmpeg}.
Utilizing the same example of \ch{C3Ti4W} we produce the visualization data with the following command.
\begin{lstlisting}[style=aflow,language=aflowBash]
aflow --soliquidy --render --auid=03cb24fa8c754e9d
\end{lstlisting}
Next we create an image with \texttt{tachyon}, and  execute the automatically generated \texttt{bash} script \texttt{render.sh} to produce an animation for carbon.
\begin{lstlisting}[style=aflow,language=aflowBash]
cd ./03cb24fa8c754e9d/
tachyon .Ti.ty -o Ti.ppm -format PPM -auto_skylight 0.7 -fullshade -numthreads 8
cd ./animation_C && bash render.sh
\end{lstlisting}

\asubsection{Dielectric function}
Rapidly growing fields such as plasmonics~\cite{Murray_ADVMAT_2007}, metamaterials~\cite{Zheludev_NMAT_2012}, and nanophotonics~\cite{Koenderink_SCIENCE_2015} leverage the interaction between light and materials.
These interactions are encoded in the material’s dielectric function, which, though costly, can be calculated directly from first-principles~\cite{Hybertsen_PRB_1987}.
Thus, we have introduced the calculation of the dielectric function as a new calculation type in \AFLOWF.

\begin{figure}[ht!]
  \includegraphics[width=0.5\textwidth]{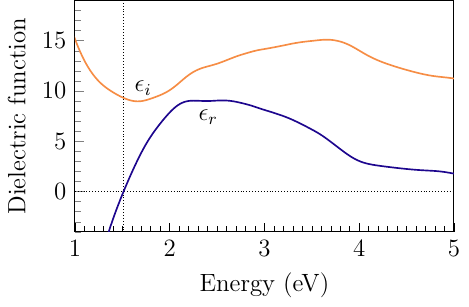}
  \caption{\small \textbf{Dielectric function} of \ch{HfNbTaTiZrC5} calculated using our workflow.}
  \label{fig:dielectric}
\end{figure}

\AFLOW\ has had three run schemes which utilize \VASP\ for ab-initio calculations: \texttt{RELAX}, \texttt{STATIC} and \texttt{BANDS}; they are responsible for geometry optimization, well-converged total energy with a dense \textbf{k}-grid, and high-resolution band structure, respectively.
The \texttt{BANDS} run uses the \texttt{STATIC} run as a basis for further calculations, and likewise, the new \texttt{DIELECTRIC} run follows the same procedure.

A \texttt{DIELECTRIC} run corresponds to a \VASP\ calculation, where the dielectric tensor is calculated in the independent particle approximation using the Green-Kubo approach~\cite{Gajdos2006}.
To make sure that there are no spurious oscillations in the dielectric function, we use a \textbf{k}-grid that is twice as dense as the one used for \texttt{STATIC} runs.
The \VASP\ input and output filenames are appended by \texttt{.dielectric} to differentiate them from the other run types.
The real ($\epsilon_r$) and imaginary ($\epsilon_i$) part of the dielectric function for a high-entropy carbide, calculated with this workflow {using the partial occupation method (see the subsequent section)}, is shown in Fig.~\ref{fig:dielectric}.

Finally, as before, it is possible to incorporate multiple run schemes in one workflow.
For example, \texttt{RELAX\_STATIC\_DIELECTRIC} will perform the following runs in order: \texttt{RELAX}, \texttt{STATIC} and
\texttt{DIELECTRIC}, without any additional user input.
{Systems in the database with the \texttt{DIELECTRIC} run contain two new keywords \texttt{freq\_plasma} and \texttt{dielectric\_static}, corresponding to the bulk plasma frequency and the static dielectric tensor, respectively.}

\asubsection{Machine-readable export}
First-principles software generates a substantial amount of data that is often not machine-readable and may not be directly useful to the user.
This is of paramount importance in the context of feature selection and processing for training machine-learning models, which are ubiquitous in materials science~\cite{Morgan_ARMR_2020} and can be used to calculate exotic properties, like spinodal decomposition~\cite{curtarolo:art201}.

In the new version, \AFLOW\ can process multiple completed runs to create a single \JSON\ output which contains the lattice vectors, total energies, forces, stresses and positions of the ions for each ionic relaxation step.
This functionality is invoked by:
\begin{lstlisting}[style=aflow,language=aflowBash]
aflow --export=iap --alloy=ALLOY [--aflowlibpath=PATH]
\end{lstlisting}
which will search through the user’s \AFLOW\ database for systems that contain exactly the elements that are found in the \texttt{alloy} field.
The user can also specify the database path with the \texttt{aflowlibpath} option, if the files are stored in a non-standard location or if the user wants to only target specific runs.

This new functionality will accelerate designing machine-learning potentials by circumventing the need to write an interface to read the output files.
In the future, we plan to extend our \JSON\ export functionality to cover many more features generated by \AFLOW.

\begin{figure}[ht!]
  \includegraphics[width=0.5\textwidth]{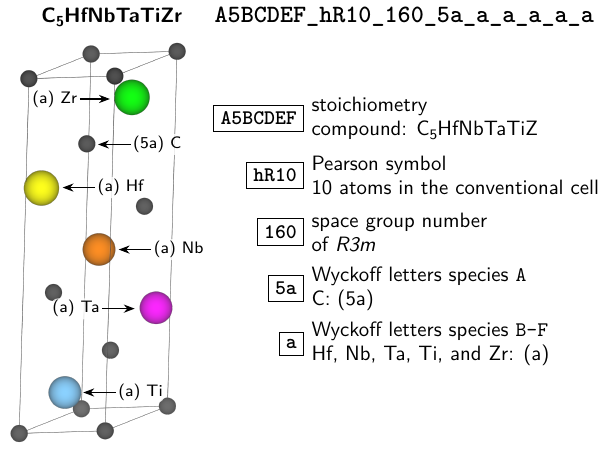}
  \caption{\small \textbf{An example of a \POCC\ tile for \ch{HfNbTaTiZrC5}} with its corresponding prototype label.}
  \label{fig:proto}
\end{figure}

\begin{figure*}[ht!]
    \centering
    \includegraphics[width=0.99\textwidth]{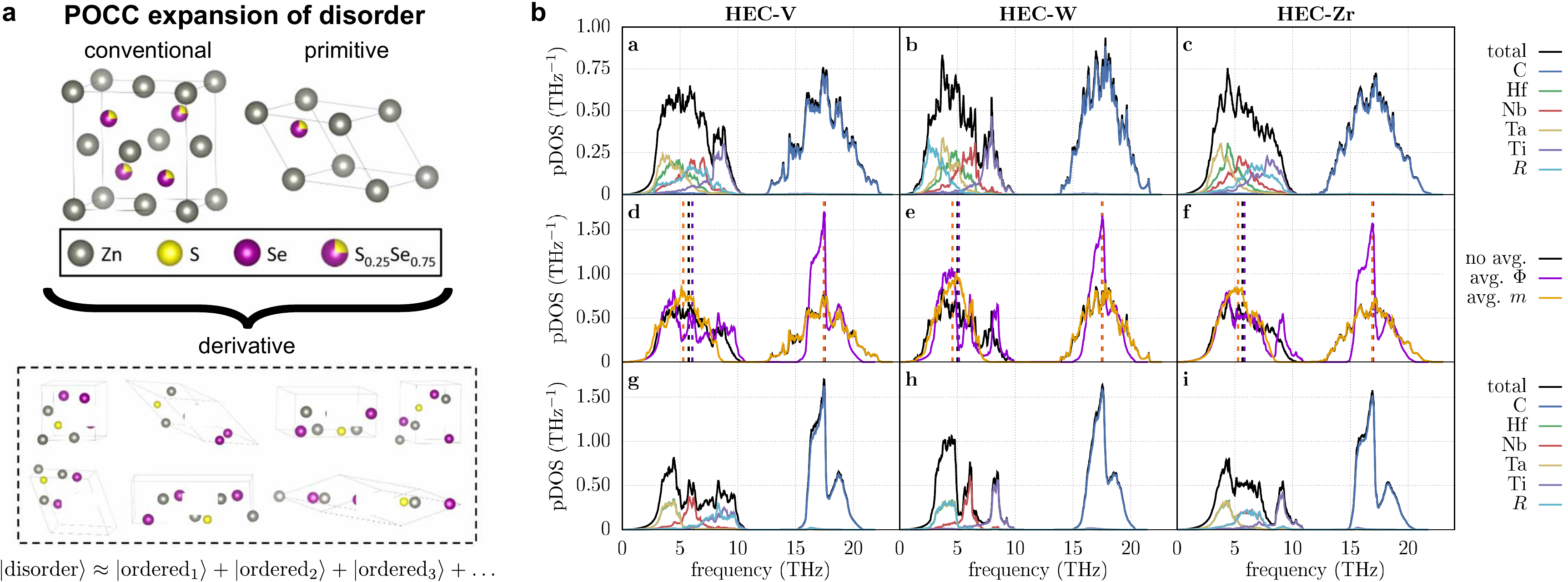}
    \caption{\small
    {\bf POCC ensemble expansion.}
    \textbf{a} Illustration of how POCC expands a disordered structure into an ensemble of derivative ordered structures, called POCC tiles.
    \textbf{b} Phonon densities of states (pDOS) computed from the POCC method.
    Adapted from Ref.~\onlinecite{curtarolo:art180} (CC BY 4.0).}
    \label{fig:pocc}
\end{figure*}

\asubsection{Prototypes}
Crystal prototypes play a pivotal role in organizing and accessing the vast array of possible crystal structures, which are crucial for advancing materials science research.
The \AFLOW\ Prototype Label offers a systematic approach to encoding key structural information, such as stoichiometry, Pearson symbol, space group, and Wyckoff positions~\cite{curtarolo:art121,curtarolo:art145,curtarolo:art173,curtarolo:art210}.
An example is presented in Fig.~\ref{fig:proto}, showing the usage of prototypes to generate tiles for the partial occupation method.
This enables efficient classification and retrieval of crystal data, forming the core of the \AFLOW\ framework, as we utilize these prototypes for rapid high-throughput materials discovery.
The associated Library of Crystallographic Prototypes\footnote{\href{https://aflow.org/prototype-encyclopedia/}{https://aflow.org/prototype-encyclopedia/}} has significantly expanded from 298~\cite{curtarolo:art121} to over 2000 entries~\cite{curtarolo:art210}.
This dynamic resource is continuously updated and enhanced, including from user contributions, to ensure its relevance.
All additions are curated from current literature to reflect the evolving research landscape and to guarantee accuracy.
Overall, these prototypes are instrumental in accelerating materials innovation by facilitating quick structural searches and predictions.

To find an arbitrary structure's \AFLOW\ prototype label, use the following command line:
\begin{lstlisting}[style=aflow,language=aflowBash]
aflow --prototype < STRUCT_FILE
\end{lstlisting}
In the reverse case, a structure input file can be generated from a prototype label and a set of parameters:
\begin{lstlisting}[style=aflow,language=aflowBash]
aflow --proto=A5BCDEF_hR10_160_5a_a_a_a_a_a --params=3.7093,12.2474,0.0,0.6,0.2,0.8, 0.4,0.3,0.9,0.5,0.1,0.7 > POSCAR
\end{lstlisting}
Both functions took over six times longer in the old \AFLOW\ version.
Although the individual time savings may be only a few seconds, these cumulative time savings become significant when preparing high-throughput studies involving hundreds of thousands of structures.

\begin{figure*}[ht!]
    \centering
    \includegraphics[width=0.9\textwidth]{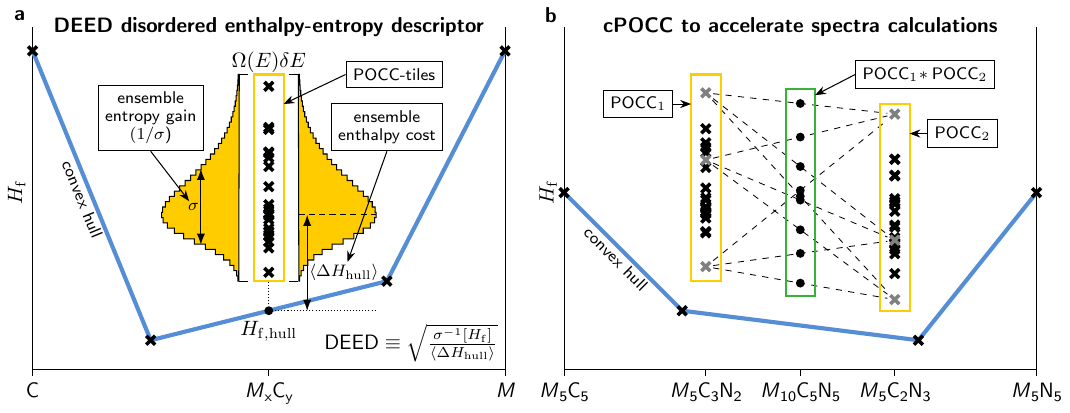}
    \caption{\small
    {\bf DEED and cPOCC workflows. }
    \textbf{a}, DEED workflow. Calculations of the phase diagram convex hull (blue line) and POCC-tiles (black crosses) give enthalpy spectral momenta. During synthesis, the enthalpy ensemble average is associated to the enthalpy loss of order, while the variance gives the entropic disorder gain. DEED is the balance of the two.
    \textbf{b}, cPOCC workflow. The POCC expansion of a \ch{\textit{M}10C5N5} (\ch{\textit{M}} = metal species) high-entropy ceramic (green box) with five unique metal species would require 17.5 million 20-atom tiles. This is overcome by a partition into \ch{\textit{M}5C2N3} and \ch{\textit{M}5C3N2} subsystems (yellow boxes), each giving 490 ten-atom tiles (black crosses) and the subsequent energy convolutions (black circles). Adapted from Ref.~\onlinecite{curtarolo:art200} (CC BY 4.0).}
    \label{fig:deed}
\end{figure*}

\asubsection{Partial occupation method}
The inherent chemical disorder in high-entropy materials has been a long-standing challenge for the accurate modeling and prediction of materials properties.
There are two primarily approches to model this disorder: supercell or spectral.
Supercell methods represent the disordered system by large, ordered configurations of atoms.
However, they cannot extract any finite-temperature information about the disordered material, such as phase transitions.
In contrast, spectral methods use a Boltzmann average of small, ordered configurations.
Therefore, they are more suitable for investigating the phase diagram at arbitrary temperatures.

The partial occupation (\POCC) method belongs to the spectral category.
It models materials by employing Boltzmann statistics to an ensemble of structure calculations, resolving properties of disordered systems~\cite{aflowPOCC}.
The implementation of the \POCC\ module provides an efficient and convenient method for the prediction of useful properties of disordered systems.

The workflow of the \POCC\ algorithm, illustrated in Fig.~\ref{fig:pocc}a, is as follows:
(i) for a given prototype with fractional decorations, generate a set of unique cells with appropriate stoichiometries (tiles) as the one shown in Fig.~\ref{fig:proto};
(ii) utilizing the \AFLOW\ framework, calculate the properties of the tiles using \DFT;
(iii) perform Boltzmann statistics for the property of interest, as a function of temperature, using the tiles formation enthalpies.
For example, in Fig.~\ref{fig:pocc}b, the phonon density of states is calculated for a high-entropy system using the \POCC\ method.

{This method is not limited to the averaging of scalar quantities, but can also be used to average tensorial quantities, such as the dielectric function.
As the other properties, the dielectric function of each tile is computed and then an ensemble average is performed to calculate the dielectric function of the disordered system, as was done in Ref.~\citenum{curtarolo:art187} or the example given in Fig.~\ref{fig:dielectric}.
Note that the dielectric tensor for cubic systems, like the high-entropy carbides, is isotropic, so there is only one independent component~\cite{nye_symmetry}.}

The disordered system is specified either using the \AFLOW\ prototype label or a \texttt{PARTCAR} file.
This file is similar in structure to a \VASP\ \texttt{POSCAR} file, allowing stoichiometries to be specified on a site-by-site basis.

Despite the advantages introduced by \POCC, it can still become prohibitively expensive to model multi-component disordered systems at certain stoichiometries.
This is addressed by our convolutional algorithm \CPOCC, which allows a \POCC\ system to be described by a convolution of subsystems with complementary stoichiometries.
This is exemplified in the success of calculating properties of a five-metal carbonitride in which \CPOCC\ reduced the cost by several orders of magnitude~\cite{curtarolo:art200}.

The \texttt{PARTCAR} for a \POCC\ system can be generated from a slightly modified prototype label where occupancies are specified per species:
\begin{lstlisting}[style=aflow,language=aflowBash]
 aflow --proto=AB_cF8_225_a_b:C:Hf:Nb:Ti:Ta:Zr --pocc_params=S0-1.0xA_S1-0.2xB-0.2xC-0.2xD-0.2xE-0.2xF > PARTCAR
\end{lstlisting}
or, specified by positions:
\begin{lstlisting}[style=aflow,language=aflowBash]
 aflow --proto=AB_cF8_225_a_b:C:Hf:Nb:Ti:Ta:Zr --pocc_params=P0-1.0xA_P1-0.2xB-0.2xC-0.2xD-0.2xE-0.2xF > PARTCAR
\end{lstlisting}
In either case, the indices \texttt{Pi} and \texttt{Si} refer to the respective index of the position and species as they would appear in the parent structure, i.e. that which would be generated by:
\begin{lstlisting}[style=aflow,language=aflowBash]
 aflow --proto=AB_cF8_225_a_b:A:B:C:D:E:F
\end{lstlisting}

The number of unique supercells can also be determined from a \texttt{PARTCAR}:
\begin{lstlisting}[style=aflow,language=aflowBash]
 aflow --proto=AB_cF8_225_a_b:C:Hf:Nb:Ti:Ta:Zr --pocc_params=S0-1.0xA_S1-0.2xB-0.2xC-0.2xD-0.2xE-0.2xF | aflow --pocc_count_unique_fast
\end{lstlisting}

\asubsection{Convex hull}
The \CHULL\ module~\cite{curtarolo:art144} is designed to compute convex hulls for systems with even a large number of components.
For binary and ternary hulls it can generate illustrations and comprehensive reports summarizing the data.
The module distinguishes between stable (ground-state) and unstable designations for each compound.
It also calculates key metrics such as distances to the hull, decomposition reactions, and the stability criteria.
Additionally, the report identifies the phases that can coexist in equilibrium for each stable phase.

Furthermore, \CHULL\ can calculate the N+1 enthalpy gain, which quantifies how much a compound with N-component deviates from the convex hull constructed using compounds with fewer species (\{1, $\cdots$, N--1\}).
For 1-component systems, this value corresponds to the cohesive energy, while for 2-component systems, it represents the formation enthalpy.

An analysis of systems only containing metal species reveals that the enthalpy gain decreases as the number of species increases.
However, this trend is counteracted by the configurational entropy gain when there are at least four components (N $\geq$ 4).
This observation highlights the inevitable presence of disorder in multi-component systems~\cite{curtarolo:art152}.

Recently, we have introduced the disordered enthalpy-entropy descriptor (\DEED)~\cite{curtarolo:art200}, which captures the functional synthesizability, the likelihood of successful formation of a material under specific conditions.
The workflow for \DEED\ is shown in Fig.~\ref{fig:deed}.
This descriptor uses the distance from the convex hull to quantify the enthalpy costs that balances the entropy gains.
Providing data on how far phases are from stability helps predict the ease of synthesis, allowing accurate material prediction.

The CHULL module offers multiple interfaces: it can be accessed via the web~\cite{curtarolo:art190}, locally with the \AFLOW\ binary, or invoked through \texttt{python}.
The \AFLOW\ binary provides full access to all functionalities and options, including various output formats such as plain text, JSON, PDF.
Note that generating PDF outputs requires the \LaTeX\ package.
The \CHULL\ functionality can be invoked using the following command:
\begin{lstlisting}[style=aflow,language=aflowBash]
aflow --chull --alloy=ALLOY
\end{lstlisting}
The optimization of the various aspects of \AFLOW\ significantly improved the creation of convex hulls.
For example, the execution time for the \ch{MnPdPt} alloy was 3.4 times faster.

\asubsection{Coordination corrected enthalpies}
The coordination corrected enthalpies (\CCE) method addresses the challenge of accurately predicting formation enthalpies, especially for ionic systems where standard density functional theory (\DFT) approaches often introduce substantial errors well above the room temperature energy scale~\cite{curtarolo:art150}.
Unlike earlier correction schemes that relied solely on material composition (stoichiometry), \CCE\ incorporates structural information by parameterizing \DFT\ errors based on bond counts and oxidation states.
This approach enhances accuracy and transferability by focusing on topological bonding characteristics, which are fundamental to materials properties.
Validation across 70+ ternary oxides~\cite{curtarolo:art150} and nitrides~\cite{Friedrich_JCP_2024} demonstrated a four- to seven-fold reduction in mean absolute errors for \PBE~\cite{PBE}, \LDA~\cite{DFT,von_Barth_JPCSS_LSDA_1972}, and \SCAN~\cite{Perdew_SCAN_PRL_2015}.
Moreover, \CCE\ successfully improves the relative stability of polymorphs~\cite{curtarolo:art150}.
The implementation of \CCE\ realizes automatic corrections for diverse systems, including 2D materials, peroxides, and mixed-valence compounds, making it a valuable enabler for materials design projects~\cite{curtarolo:art172}.

\CCE\ can thus be employed to correct convex hull phase diagrams of ionic materials systems and also to properly renormalize the enthalpic distance of the POCC ensemble -- representing high-entropy materials -- to the hull. This is crucial for \DEED~\cite{curtarolo:art200} which makes use of such relative enthalpies to predict synthesizability of high-entropy ceramics.

To print the comprehensive analysis, including cation coordination numbers, oxidation numbers, \CCE\ corrections, and formation enthalpies for a given structure file, execute the following command:
\begin{lstlisting}[style=aflow,language=aflowBash]
aflow --cce=STRUCT_FILE
\end{lstlisting}
Our code enhancements enabled the calculation of \CCE\ corrections to be performed 7.4 times faster.

\section*{AFLOW4 development update}
\AFLOW4\ generates extensive data from diverse sources, particularly for complex systems with numerous species, which poses significant computational challenges.
To overcome these challenges, we refactored the existing code to capitalize on the advanced features of \texttt{C++17}.
We have made substantial improvements to almost every aspect of the ecosystem, including installation and documentation, with portability in mind.
This ensures that \AFLOW\ operates efficiently on various computer systems.

\asubsection{Build process}
The first change that will be noticed by users of \AFLOW\ is the new tool we selected to build our executable.
While using a \texttt{Makefile} worked well in the past, we wanted to enable more flexibility in the build process of \AFLOW.
Therefore, \AFLOW's build-system for version 4 is based on \texttt{cmake}.
While the switch also resulted in a 24\% speedup in build time for eight CPU cores, the primary reason for choosing \texttt{cmake} is that it allows us to efficiently provide various build presets for \AFLOW\ users.
Users can now easily select whether they want \AFLOW\ to be linked dynamically or statically.
This became a more important consideration as we significantly reduced the usage of external binaries and transitioned to directly using the corresponding libraries, such as \texttt{libarchive} or \texttt{OpenSSL}.
The option to build a version of \AFLOW\ that includes these essential libraries directly within the executable is tailored for \HPC\ systems, where consistently providing libraries on every computational node can be complex, and installing new system-wide packages is often tightly regulated.
To compile the correct versions of the static libraries, we used the integration of \texttt{cmake}  with the open source dependency manager \texttt{vcpkg}.
In comparison, the linked version of \AFLOW\ enables a smaller memory footprint and faster build times, perfect for personal computers.

Overall, \AFLOW\ is designed to run on Unix-like systems and is actively developed on \texttt{Linux} and \texttt{macOS}.
While it does not natively support \texttt{Windows}, it is still possible to use \AFLOW\ through the Windows Subsystem for Linux.
A detailed description of dependencies and the build process, including the usage of \texttt{cmake} and \texttt{vcpkg}, can be found in the \texttt{README.md}\footnote{\href{https://github.com/aflow-org/AFLOW/blob/release/README.md}{https://github.com/aflow-org/AFLOW/blob/release/README.md}} file at the root of the source distribution.

\asubsection{Documentation}
High-quality documentation is crucial for the reproducibility, maintainability, and collaborative potential of the \AFLOW\ codebase.
It serves as a bridge between our technical implementation and scientific understanding.
While \AFLOW\ has always provided general documentation that offers a broad overview of the code’s functionality and workflows, code-level documentation that focuses on specific modules, functions, and algorithms was not prioritized.
Both types of documentation are essential to lower the barrier to entry for new users and contributors, enabling fellow scientists and students to understand and extend the codebase more effectively.

To address this issue, we implemented a multi-step approach to improve the documentation of our codebase.
First, we established a standardized documentation style using Doxygen, ensuring consistency across the entire codebase.

Second, we made the creation of detailed comments following this standard mandatory for all new code submissions.
Thirdly, we gradually expanded this practice to legacy code, an ongoing process due to the challenge posed by over half a million lines of existing code.
Lastly, the code level is automatically compiled into an easy-to-use website, allowing quick access to this information\footnote{\href{https://aflow.org/aflow-documentation}{https://aflow.org/aflow-documentation}}.
We hope that prioritizing clear and accessible documentation will help inspire external contributions to \AFLOW\ in the future.

\asubsection{Version control and unit testing}
With the release of \AFLOWF, we are also making significant changes to how we publish new versions of \AFLOW.
Transitioning from compressed release archives to hosting the code on GitHub\footnote{\href{https://github.com/aflow-org/AFLOW}{https://github.com/aflow-org/AFLOW}} represents a significant advancement in fostering collaboration, transparency, and code quality.
By leveraging GitHub's collaborative platform, users are empowered to provide immediate feedback, report issues, and propose enhancements or extensions to the software.
This direct engagement will help us quickly identify and fix bugs, leading to a more robust \AFLOW\ in the long run.

Automated testing integration, through an extensive new unit test suite, is another key enhancement accompanying this migration.
Similar to our updated approach to documentation, unit tests for code that is newly introduced to \AFLOW\ are mandatory.
The number of tests for legacy code is also being constantly expanded.
This automated testing ensures that every proposed code change undergoes rigorous testing before merging into the primary branch and being released.
This systematic approach to quality assurance ensures the integrity of the \AFLOW\ codebase.

\asubsection{Overall code improvements}
In this iteration of \AFLOW, we again implemented several significant enhancements to improve its performance and functionality.
Firstly, we improved communication between \AFLOW\ and external processes, streamlining data exchange and, therefore, avoiding bottlenecks in parallel computations.
Secondly, we eliminated the dependency on the separate \texttt{aflow-data} executable by embedding all necessary data directly into the binary in a compressed form.
This data is decompressed on-demand, simplifying the workflow and reducing \AFLOW's storage and memory footprint.
Thirdly, we underwent a significant restructuring of the code base, enabling us to take advantage of more aggressive compiler optimizations.
Organizing the code in a manner that aligns better with modern compiler technologies allows it to utilize contemporary computing architecture more effectively.
This optimization resulted in significant speedups across various functionalities of \AFLOW.
Lastly, we placed a strong emphasis on multi-threaded programming techniques, enabling \AFLOW\ to better utilize the high core counts found in today's \HPC\ systems.
These improvements collectively enhance the code's efficiency, scalability, and usability, ensuring that \AFLOW\ remains a valuable tool at the forefront of computational material science research.

\begin{figure}
    \centering
    \includegraphics[width=0.5\textwidth]{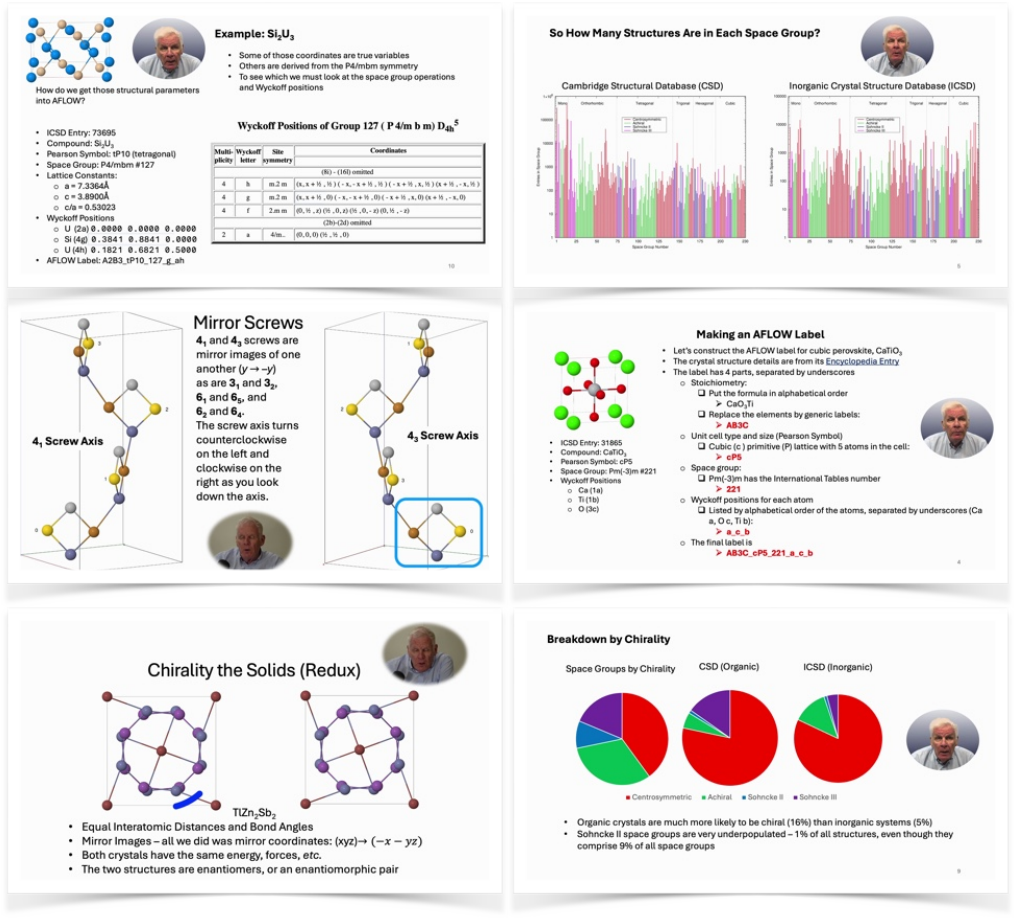}
    \caption{\small
    \textbf{\AFLOW\ tutorial videos}. Collection of screenshots from the first set of tutorials.}
    \label{fig:tutorial}
\end{figure}

\asubsection{Education}
A thriving scientific ecosystem must be educationally sustainable, providing young researchers with the opportunities and resources necessary to acquire essential skills.
This foundation enables them to contribute effectively to their field and drive innovation.
Without such educational support, new researchers may struggle to utilize available tools, potentially hindering progress significantly.

To address this, we are developing a comprehensive series of \AFLOW\ tutorials\footnote{\href{https://aflow.org/prototype-encyclopedia/tutorials.html}{https://aflow.org/prototype-encyclopedia/tutorials.html}} (Fig.~\ref{fig:tutorial}). These tutorials, available in both video and text formats, capture the fundamental concepts behind the ideas and algorithms used in the various modules of \AFLOW.
This dual format enhances accessibility, catering to different learning preferences.
Beyond individual benefits, these tutorials serve as a knowledge-preservation tool, ensuring that underlying principles are not lost over time.
They foster a robust, sustainable academic community where understanding the algorithms' core ideas leads to higher-quality research and encourages collaboration.

In tandem with the updated \AFLOW\ documentation strategy, the \AFLOW\ tutorials empower researchers to build upon \AFLOW\ effectively by providing clear examples and practical applications.
This educational initiative will strengthen the \AFLOW\ ecosystem in the years to come.

\section*{Conclusion}
\label{sec:conclusion}
Over the past few years, we have been actively optimizing and developing original modules in \AFLOW\ to facilitate the discovery of novel chemically and structurally disordered high-entropy materials.
In addition to these developments, we have made significant updates to almost all aspects of the toolkit, which is now called \AFLOWF.
These updates include enhancing the user experience by providing more user-friendly instructions on how to obtain and install the code, as well as improving the documentation to make it more accessible and understandable.
Furthermore, we’ve prioritized portability, ensuring that \AFLOWF\ runs efficiently on a wide range of computer systems.
We hope that current and future users will fully utilize the capabilities of \AFLOWF\ to advance the field of high-entropy disordered materials.

{\small
\aasubsection{Author Contributions}
Authors SD and HE contributed equally and lead the project on the modernization.
SD, HE, ST, SG, RF, NHA, and DH wrote and updated the \AFLOWF\ code.
SC supervised the project.
All authors, SD, HE, ST, SG, RF, MJM, NHA, DH, ME, NH, XC, AC, and, SC contributed to the discussion and writing of the article.
All authors have given approval to the final version of the manuscript.

\aasubsection{Acknowledgments and Funding}
The authors thank
Cormac Toher, Paolo Colombo, Art Fortini, Ohad Levy, and Amir Natan for fruitful discussions.
This research was supported by the Office of Naval Research under grant N00014-24-1-2502, N00014-23-1-2615 and N00014-24-1-2768.
This work was supported by high-performance computer time and resources from the DoD High Performance Computing Modernization Program (Frontier).
We acknowledge Auro Scientific, LLC for computational support.

\aasubsection{Data availability}
The data employed to generate the predictions are freely available from the \AFLOWorg\ database.
Processed data used in this project are available from the corresponding author upon reasonable request.
The underlying software used in this study and its source code can be accessed via this link \url{https://aflow.org/install-aflow/}.
}

\section*{Declarations}
{\small
\aasubsection{Conflict of interest}
The authors declare that they have no competing financial interests
or personal relationships that could have appeared to influence the work
reported in this article.
}

\newcommand{\Ozolins}{Ozoli{\c{n}}{\v{s}}}

\end{document}